\documentclass[showpacs,jcp,12pt]{revtex4-1}
\usepackage[english]{babel}
\usepackage{amsmath,graphicx,amssymb} 
\usepackage{times}
\usepackage{epsfig} 
\usepackage{epstopdf} 

\begin{document} 
\title{A diabatic state model for double proton transfer\\
in hydrogen bonded complexes
}

\author{Ross H. McKenzie}
\email{email: r.mckenzie@uq.edu.au}
\homepage{URL: condensedconcepts.blogspot.com}
\affiliation{School of Mathematics and Physics, University of Queensland,
  Brisbane, 4072 Queensland, Australia}

\date{\today}
                   
\begin{abstract}
Four diabatic states are used to construct a simple 
model for double proton transfer in  hydrogen bonded complexes.   
Key parameters in the model are the proton donor-acceptor separation $R$
and the ratio, $D_1/D_2 $, between the proton affinity of a donor
with one and two protons.
Depending on the values of these two parameters
the model describes four qualitatively different ground state 
potential energy surfaces, 
having zero, one, two, or four saddle points.
In the limit $D_2=D_1$ the model reduces to two decoupled hydrogen bonds.
As $R$ decreases a transition can occur from a concerted
to a sequential mechanism for double proton transfer.

\end{abstract}

\maketitle 

\section{Introduction}

A basic but important question in physical chemistry concerns a chemical reaction  that involves two steps: A to B to C.
Do the steps occur sequentially or  simultaneously, i.e., in a concerted/synchronous  manner?
Two examples of particular interest are coupled electron-proton transfer \cite{Migliore:2014} and double proton transfer.
The latter occurs in a diverse range of molecular systems involving double hydrogen bonds,
including porphycenes \cite{Waluk:2006,Waluk:2010,Gawinkowski:2012}, dimers of carboxylic acids, and DNA base pairs.
For the case of double proton transfer in the excited state of the 7-azaindole dimer
[a model for a DNA base pair] there has been some controversy about whether the process is concerted or sequential.  Kwon and Zewail \cite{Kwon:2007} argue that the weight of the evidence is for sequential transfer.
Based on potential energy surfaces from a simple analytical model for
two coupled hydrogen bonds \cite{Smedarchina:2007}
 and from computational quantum chemistry [at the level of
Density Functional Theory (DFT) based approximations] \cite{Smedarchina:2008}
it has been proposed that there can be three qualitatively different potential energy surfaces, depending on the strength of the coupling of the motion of the two protons:
(a) One transition state and two minima, as in the formic acid dimer;
(b) Two equivalent transition states, one maxima and two minima, as in the 4-bromopyrazole dimer; and
(c) Four transition states, one maxima and four minima, as in porphine.

In this paper a simple
model for double proton transfer based on four diabatic states is introduced.   
The key parameters in the model are the spatial separation $R$ of the atoms
between which the protons are transferred 
and the ratio, $D_1/D_2 $, between the proton affinity of a donor
with one and two protons.
Depending on the value of these two parameters
the  ground state 
potential energy surface has zero, one, two, or four saddle points.

\begin{figure}[htb] 
\centering 
\includegraphics[width=84mm]{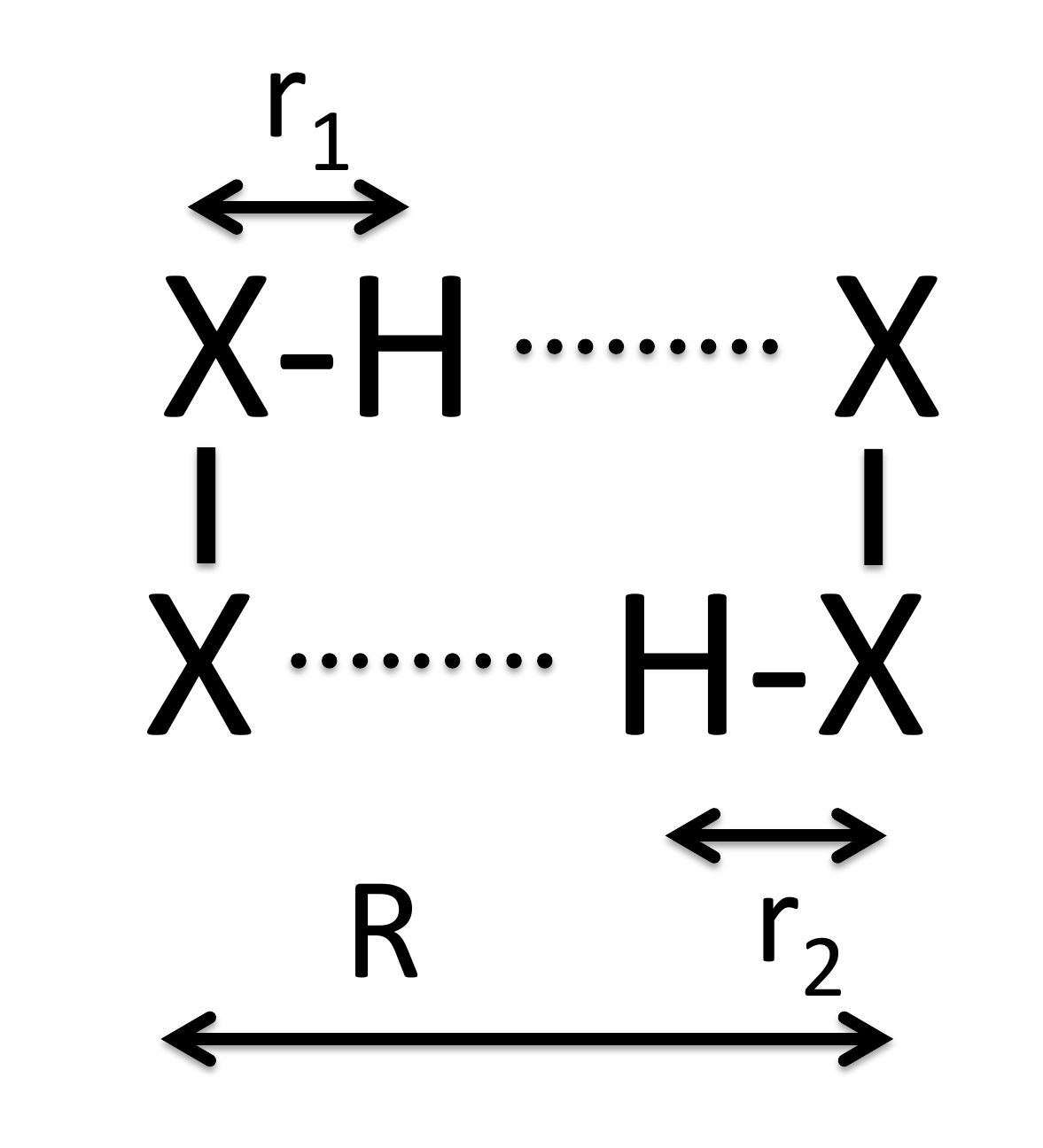}
\caption{
Definition of the key distances for a symmetrical system with two hydrogen bonds.
$R$ is the distance between the donor and acceptor atoms involved in
proton transfer. $r_1$ is the length of the upper X-H bond on the left molecule. 
 The length of the lower X-H bond on the right molecule is $r_2$. 
}
\label{fig1}
\end{figure}

\section{
A simple model for ground state potential energy surfaces
}
\label{PES}

The four-state model proposed here builds on recent work concerning single hydrogen bonds
and single proton transfer \cite{McKenzieCPL}.
A two diabatic state model with a simple parameterisation gives a ground state potential energy surface
that can describe a wide range of experimental data (bond lengths, stretching and bending vibrational frequencies, and isotope effects) for a diverse set of molecular complexes, particularly when
quantum nuclear motion is taken into account \cite{McKenzieJCP}.

\subsection{Reduced Hilbert space for the effective Hamiltonian}

\begin{figure}[htb] 
\centering 
\includegraphics[width=94mm]{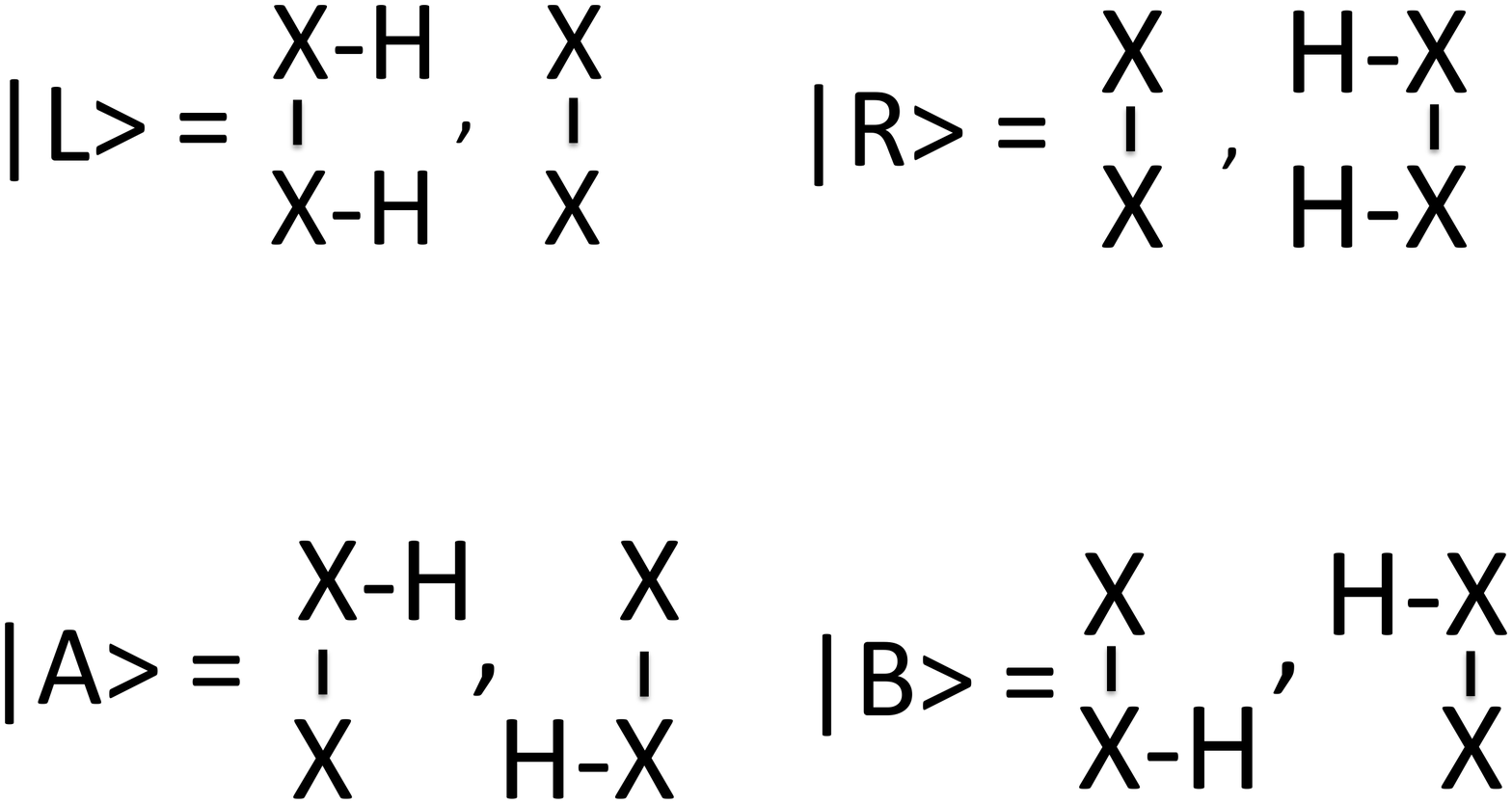}
\caption{
Four diabatic states.
The top two states differ from the bottom two by transfer
of a single proton, i.e., for the top two states both protons are on either the left molecule or the right molecule.
A key parameter in the model is $D_2/D_1$, the ratio of the proton affinity in the top two states
to that in the bottom two states.
}
\label{fig2}
\end{figure}

Diabatic states \cite{Domcke:1994,Hong:2006,Sidis:2007,VanVoorhis:2010}
(including valence bond states) are
a powerful tool for developing chemical concepts \cite{Shaik},
including understanding and describing conical intersections \cite{Yarkony:1998}, and 
going beyond the Born-Oppenheimer approximation \cite{McKemmish}.
Previously it has been proposed that hydrogen bonding
and hydrogen transfer reactions can be described by 
Empirical Valence Bond models \cite{Warshel:1991}
where the diabatic states are valence bond states.
In the model considered here, the reduced Hilbert space has a basis consisting
of the  four diabatic states  shown in Figure \ref{fig2}.
Each represents a product state of the electronic states of the left unit
and the right unit, i.e, the state that would be an eigenstate as the
distance $R \to \infty$.
The difference between the two states $|A>$, $|B>$
and the two states $|L>$, $|R>$
is transfer of a single proton.
Each of the diabatic states  involves X-H bonds; they 
have both covalent and ionic contributions, 
the relative weight of which depends on the 
length of the X-H bond.
A Morse potential is used
to describe the energy of each of these bonds and thus
the energy of the diabatic states (see below).

In this paper I focus solely on the symmetric case where
the donor and acceptor are symmetrical, i.e., 
the states $|A>$ and $|B>$ are degenerate with 
each other at their respective equilibrium bond lengths, as
are the two states $|L>$ and $|R>$.

\subsection{Effective Hamiltonian}

The   Hamiltonian for the four diabatic states
has matrix elements that depend on
the X-H bond lengths, $r_1$ and $r_2$, and the donor-acceptor separation $R$ 
(compare Figure \ref{fig1}).
For simplicity,  I assume the hydrogen bonds are linear.
It is straightforward to also take into account non-linear bonds  \cite{McKenzieCPL}.
The Morse potential describes the energy of a single X-H bond.
The two cases $j=1,2$ denote the
presence of one or two X-H bonds, respectively, within the left and right molecules shown in Figure \ref{fig2}.
The Morse potential is
\begin{equation}
V_j(r) = D_j[\exp(-2a_j(r-r_{0j})) - 2\exp(-a_j(r-r_{0j}))]
\label{morse}
\end{equation}
where $D_j$ is the binding energy, $r_{0j}$ is the equilibrium bond length,
and $a_j$ is the decay constant. 
$D_1$ and $D_2$ denote the proton affinity, with one and two protons attached, respectively.
Generally, I expect $D_2 < D_1$, and actual values for their relative size for specific molecules
are discussed later.
The harmonic vibrational frequency $\omega$
of an isolated X-H bond is given by
$\mu\omega^2=2D_j a_j^2$ where $\mu$ is the reduced mass of the proton.
For O-H bonds approximate parameters are $\omega \simeq 3750 $ cm$^{-1}$, $D \simeq 120$ kcal/mol,
$a \simeq 2.2/\AA$, $r_0 \simeq 0.96 \ \AA$.
I take $a_1=a_2$ and $r_{01}=r_{02}$.

The effective Hamiltonian describing the four interacting diabatic states
is taken to have the form
\begin{equation}
H = \left(\begin{array}{cccc} 
V_2(r_1) +   V_2(R-r_2)  & 0 & \Delta( R) & \Delta( R) \\
0 & V_2(R-r_1) +   V_2(r_2)   & \Delta( R) & \Delta( R) \\
 \Delta( R) & \Delta( R) & V_1(r_1) +   V_1(r_2)  & 0 \\
 \Delta( R) & \Delta( R) & 0 &  V_1(R-r_1) +   V_1(R-r_2) 
\end{array}\right)
\label{eqn-ham}
\end{equation}
where the basis for the four-dimensional Hilbert space is taken in the order
$|L>$, $|R>$, $|A>$, and $|B>$. 
The diabatic states that differ by one proton
in number are coupled  via the matrix element $\Delta(R)$.
The coupling between states that differ by two protons is
assumed to be negligibly small.

\subsection{Parameterisation of the diabatic coupling $\Delta(R)$}

The matrix element associated with proton transfer is assumed to have the simple form \cite{McKenzieCPL}
\begin{equation}
\Delta(R)= \Delta_1 \exp(-b (R-R_1))
\label{DeltaR}
\end{equation}
and $b$ defines the decay rate of the matrix element with
increasing $R$.
 $R_1$ is a reference distance defined as $R_1 \equiv 2r_0+1/a
\simeq 2.37 \ \AA$.
This distance is introduced so that the scale $\Delta_1 $ 
is an energy scale that is physically relevant.
There will be some variation in
the parameters $\Delta_1$ and $b$ with the chemical
identity of the atoms (e.g. O, N, S, Se, ...) in the donor
and acceptor that are directly involved in the H-bonds.  
Since the Morse potential parameters are those of isolated X-H bonds
the model for a single
hydrogen bond has essentially two free parameters, $b$, and $\Delta_1$.
These respectively set the length and energy scales associated with the interaction between
any two diabatic states that are related by transfer of a single proton.
The parameter values that are used here,
$\Delta_1=0.4D_1 \simeq$ 48 kcal/mol 
and $b=2.2/\AA$ for O-H$\cdots$O systems,
were estimated from comparisons of the predictions of the model
with experiment \cite{McKenzieCPL}, and give particularly good agreement
when quantum nuclear effects are taken into account \cite{McKenzieJCP}.
Appropriate parameter values for N-H$\cdots$N bonds
are discussed later.

\subsection{Four different classes of ground state potential energy surfaces}

\begin{figure}[htb] 
\centering
\includegraphics[width=80mm]{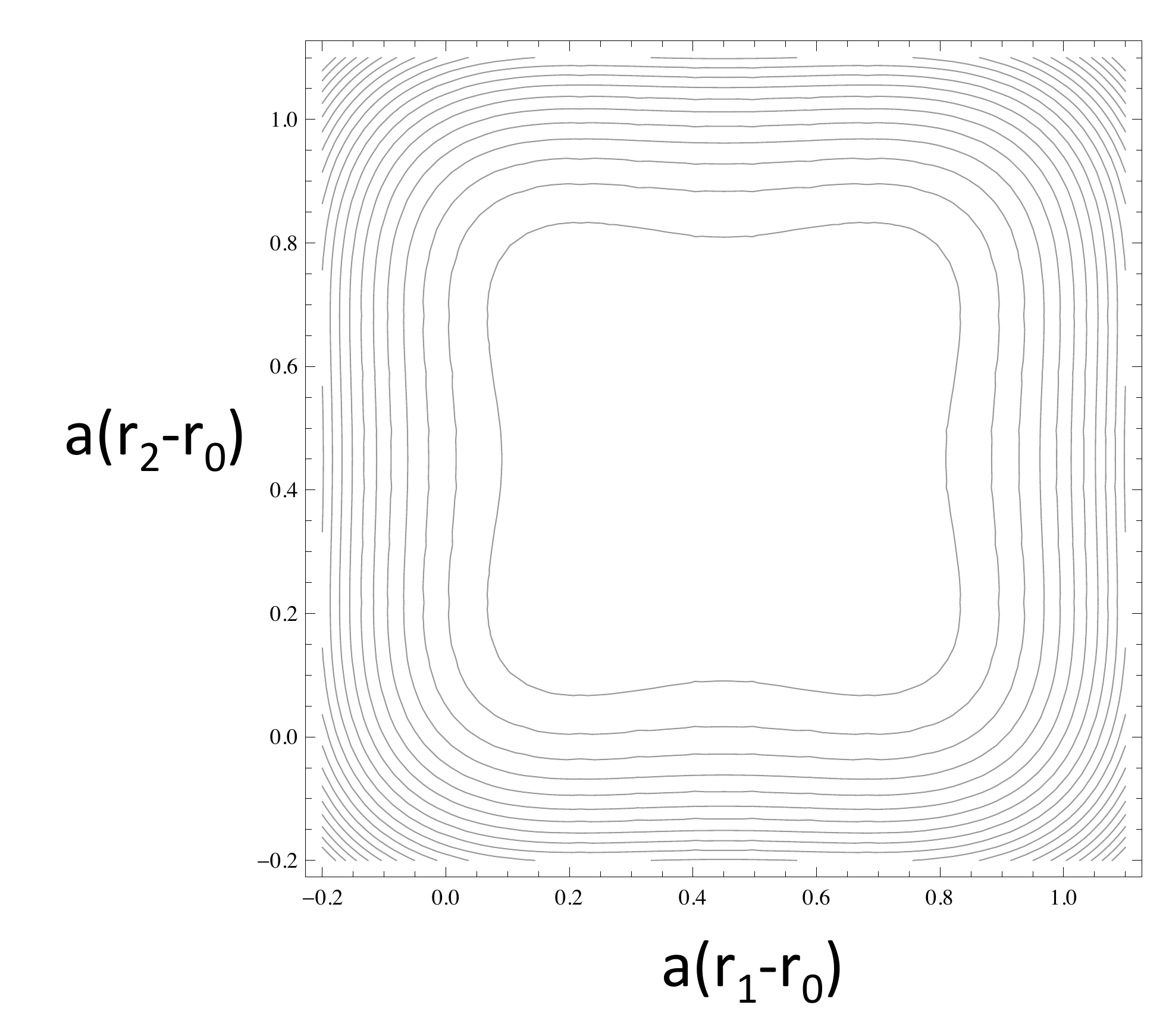}
\includegraphics[width=80mm]{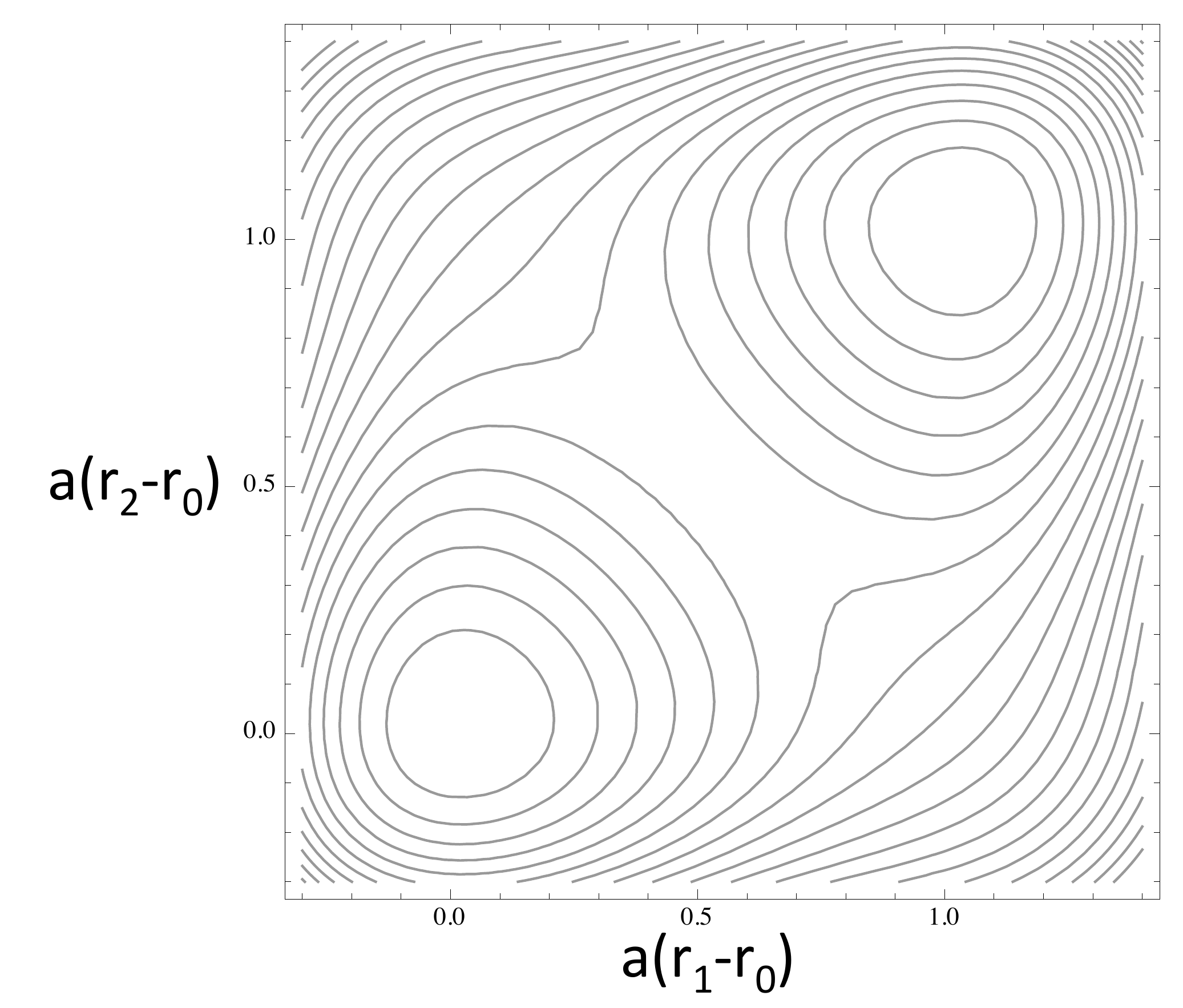}
\includegraphics[width=80mm]{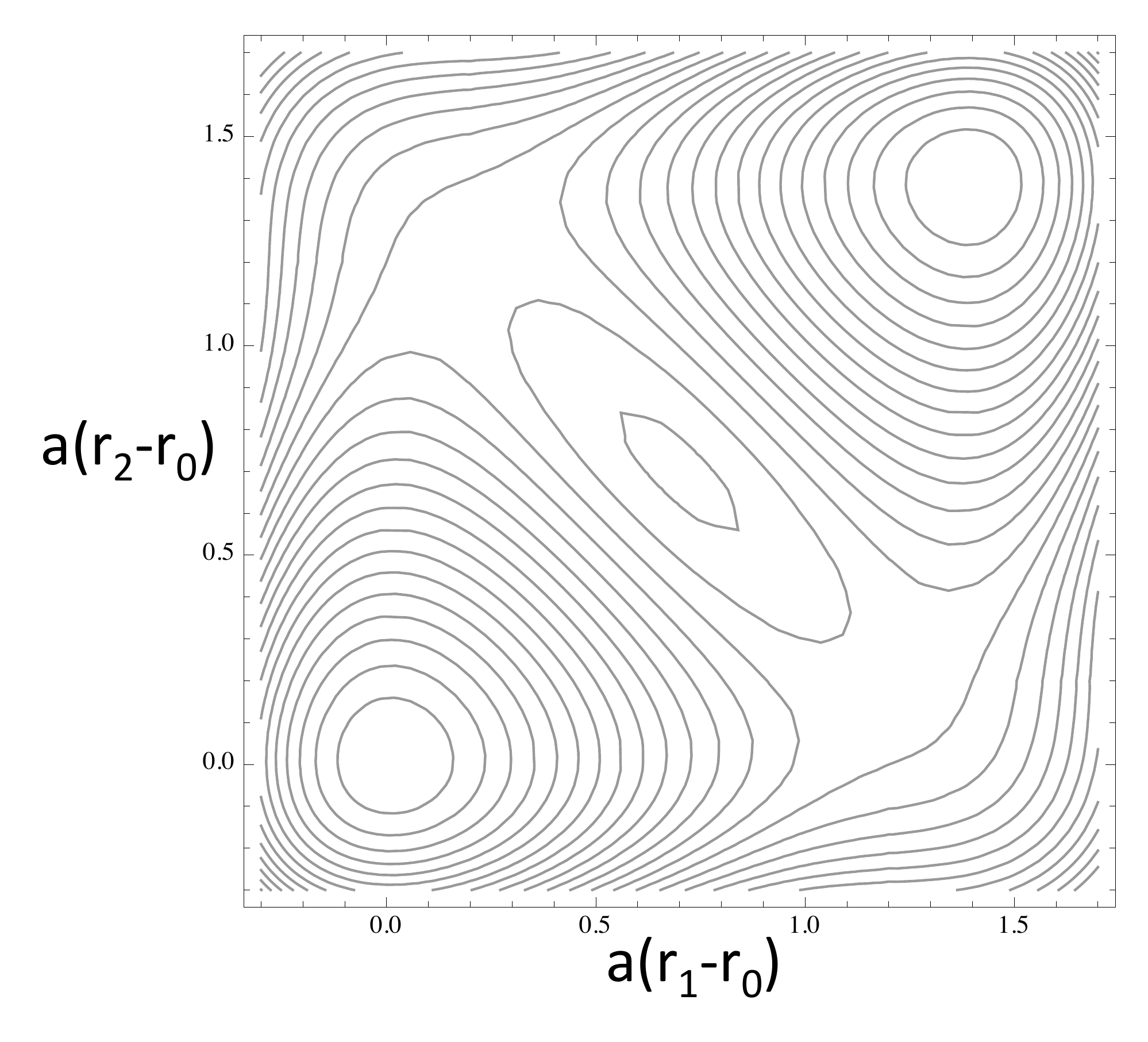}
\includegraphics[width=80mm]{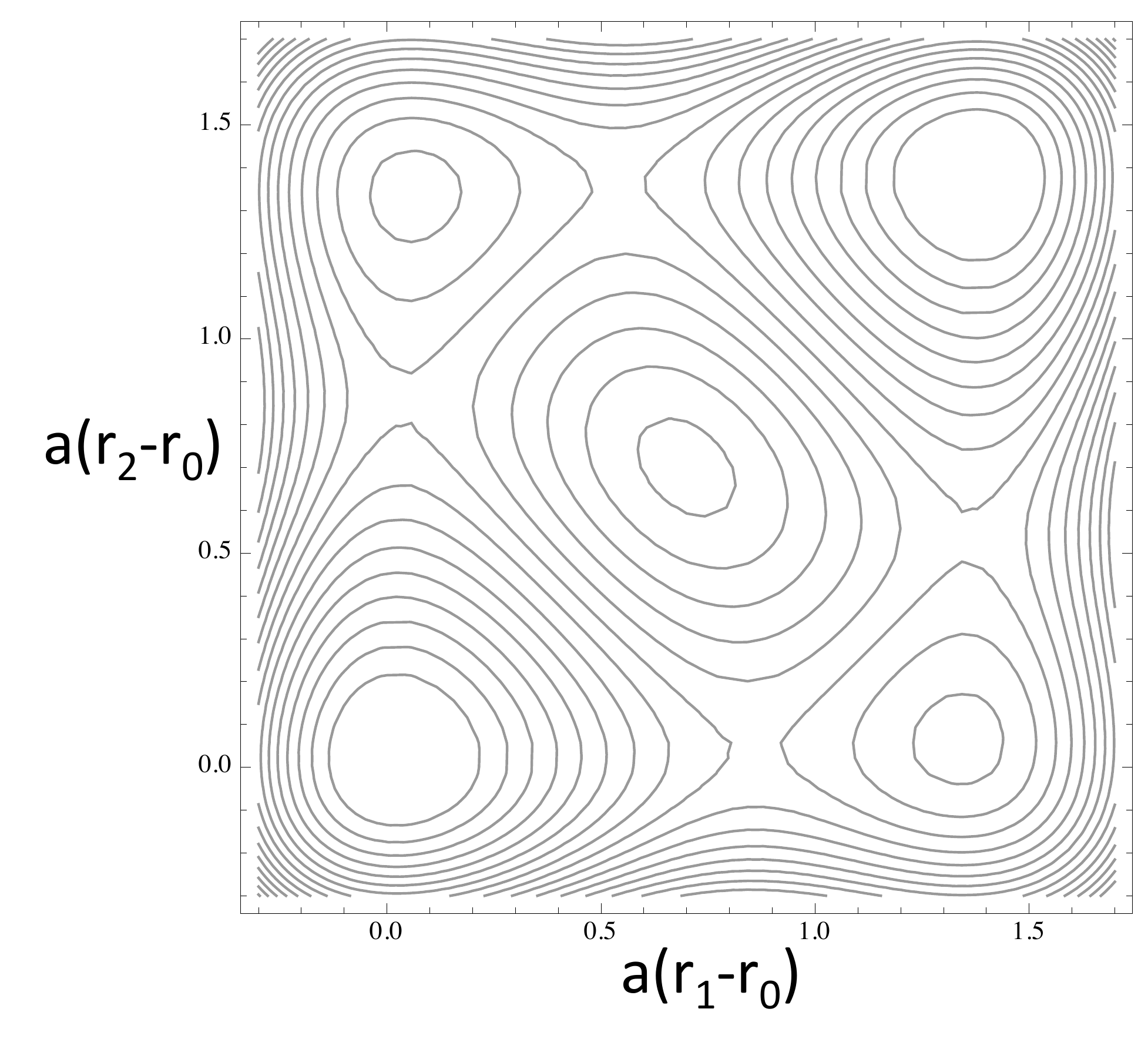}
 \caption{
Representatives of the four different classes of
two-dimensional ground state potential energy surfaces.
Top left:
Class I, no saddle point, $R=2.3 \AA,  
 D_1=D_2$.
The contour level spacing is $0.007D_1=0.84$ kcal/mol.
Top right: Class II,
 one saddle point, $R=2.4 \AA, 
D_2/D_1=0.5$.
The contour level spacing is $0.02D_1=2.4$ kcal/mol.
Bottom left: Class III,
 two saddle points, $R=2.56 \AA, 
D_2/D_1=0.7$.
The contour level spacing is $0.018D_1=2.2$ kcal/mol.
Bottom right: Class IV,
 four saddle points, $R=2.56 \AA, 
D_2/D_1=0.9$.
The contour level spacing is $0.015D_1=1.8$ kcal/mol.
Parameters relevant to O-H $\cdots$O bonds,
$r_0 = 0.96 \AA$, $a = 2.2/\AA$, and $D_1 = 120 $ kcal/mol \cite{McKenzieCPL}, were used.
}
\label{figf}
\end{figure}

In the adiabatic limit [i.e., the classical limit
where the protons are taken to have infinite mass] 
 the four electronic energy eigenvalues 
of  the Hamiltonian (\ref{eqn-ham})  define potential energy surfaces (PES)
for each  of  the four electronic states.
I focus on the smallest eigenvalue, which describes the ground state potential surface $\epsilon_0(r_1,r_2,R)$.
Figure \ref{figf} shows four qualitatively different ground state surfaces, depending on the value of
the two parameters $R$ and $D_2/D_1$. The four different classes  are now 
defined and discussed. The classes differ by the number of saddle points
on the potential surface. The last three classes were delineated previously
in Reference \onlinecite{Smedarchina:2008}.

{\it Class I.}
There is a single minimum and no local maxima on the surface.
The two protons will be completely delocalised between the four
different binding sites.

{\it Class II.}
There are two local minima and a single saddle point equidistant
between them. Double proton transfer will occur  by
the concerted mechanism. Both the minimum energy path 
(for activated transfer at high temperatures) 
and the instanton path associated with quantum tunneling
(at low temperatures) is
along the diagonal direction $r_1=r_2$.

{\it Class III.}
There are two local minima and a single maxima       equidistant
between them, and two saddle points (transition states)
on opposite sides of the maxima.
Activated transfer will occur via either of the saddle
points and thus can be described as a compromise between
concerted and sequential transfer.
Depending on the details there may be a single linear instanton path 
(concerted tunneling) or two non-linear paths on either side of
the potential maximum (partially concerted tunneling) \cite{Benderskii:2003}.

{\it Class VI.}
There are four local minima and a single maxima       equidistant
between them, and four saddle points (transition states).
Activated transfer will occur via sequential transfer.
The minimum energy path may involve a significant energy plateau
[a “structureless transition state”]
as found in some previous computational chemistry calculations for
pyrazole-guanidine \cite{Schweiger-2003}.


Classes  I, II, and III can be distinguished by examining the local curvature of
the PES at the symmetric point $r_1=r_2=R/2$. In particular
\begin{equation}
K_1 \equiv 
\left.\frac{ \partial^2 \epsilon_0(r,r,R) }{\partial r^2}\right\vert_{r=R/2}.
\label{freq1}
\end{equation}
is positive for class I and negative for all the others.
The curvature in the perpendicular direction
\begin{equation}
K_2 \equiv \left.\frac{ \partial^2 \epsilon_0(r,R-r,R) }{\partial r^2}\right\vert_{r=R/2}
\label{freq2}
\end{equation}
is positive for classes  I and II and negative for classes  III and IV.
$K_1$ and $K_2$ are proportional to the vibrational frequencies $\omega_s$ and $\omega_a$,
respectively, introduced in Ref. \onlinecite{Smedarchina:2007}, and used there to distinguish classes  II and III.

\begin{figure}[htb] 
\centering
\includegraphics[width=120mm]{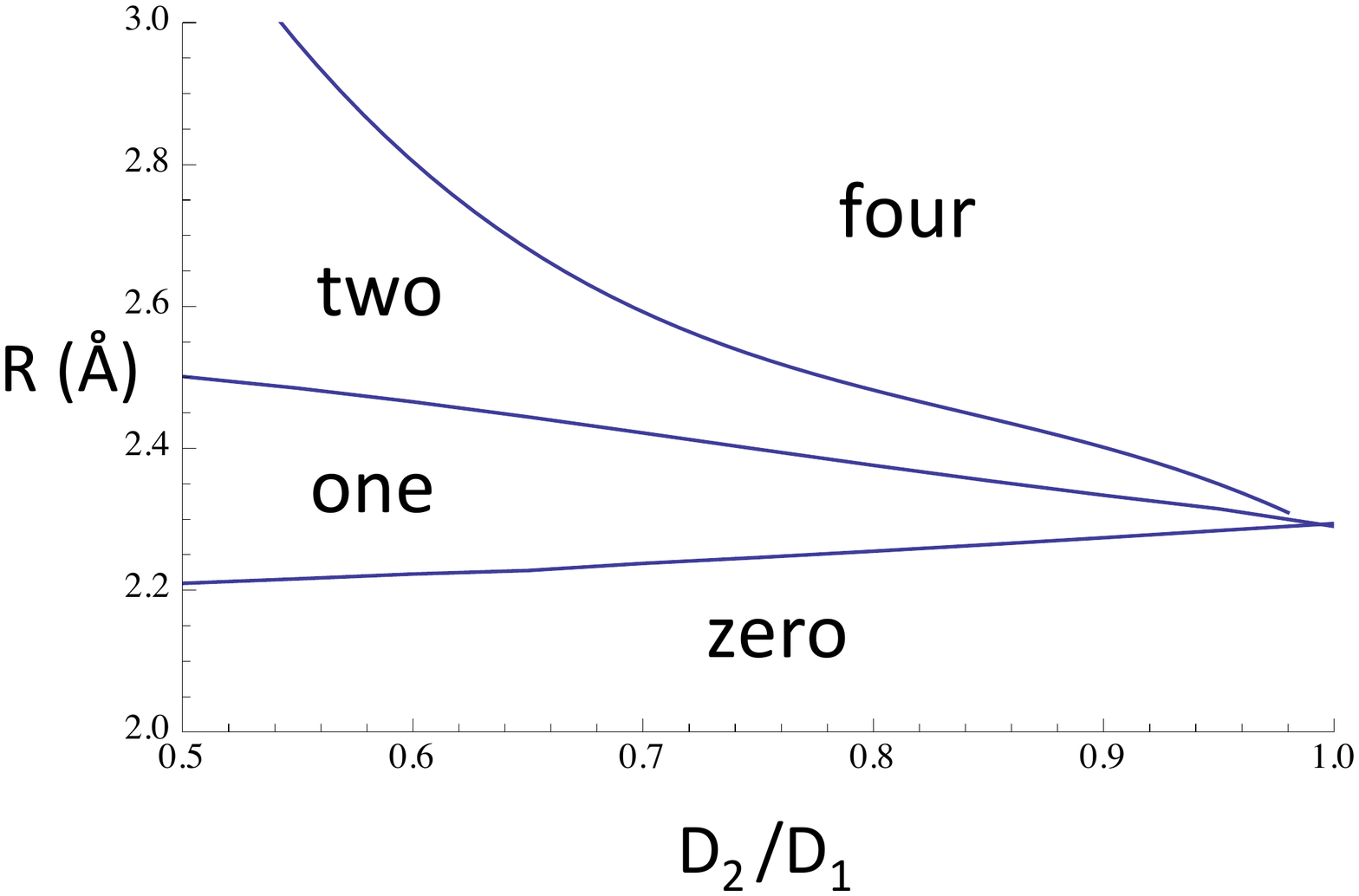}
 \caption{
Phase diagram of the model as a function of the donor-acceptor distance $R$ and 
the ratio of the double to single proton affinities $D_2/D_1$.
Parameter regions where the ground state potential energy surface 
has zero, one, two, or four saddle points are indicated. The vertical scale is for parameters
appropriate for O-H$\cdots$O bonds. The $R$ values would be larger by approximately
$0.14 \AA$ for N-H$\cdots$N bonds. 
}
\label{figg}
\end{figure}

Figure \ref{figg} shows the "phase diagram" of the system as a function of the
donor-acceptor distance $R$ and ratio of proton affinities, $D_1/D_2$, i.e., for which parameter
regions the classes  I-IV listed above occur.
In particular, for fixed $D_2/D_1$ as $R$ decreases from a large value
there will be
a transition from class IV to III to II to I.

Catastrophe theory has been used to describe the qualitative changes in energy landscapes 
that occur with variation of a system parameter \cite{Wales:2003}.
Examining the plots shown in Figure \ref{figf}, suggests that the cusp catastrophe describes 
transitions between classes I and II and between II and III. The transition between
III and IV is described by two simultaneous fold catastrophes.
 
If the extrema are isolated the relative number of minima ($N_{min}$), maxima ($N_{max}$), 
and saddle points ($N_s$) on the potential energy surface
is constrained by the relation
\begin{equation}
N_{min}+N_{max} - N_s = 1.
\label{poincare}
\end{equation}
For example, in class IV, $4+1-4=1$.
Hence, if varying the system parameters introduces an extra maxima or minima then
one additional saddle point must also appear. 
This relation is a consequence of 
differential topology [Morse theory and the Poincare-Hopf index theorem]. The minima and maxima are associated with an
index $+1$ and saddle points with $-1$, where the index is $(-1)^J$ with $J$ the number of negative eigenvalues of the Hessian matrix at the extremal point.
 A general theorem \cite{McKenzie83} states
that if a smooth function $f(\vec{r}) \to \infty$ as $|\vec{r}| \to \infty$ or if the gradient of $f$ points outward
over a closed surface (curve in two dimensions), then the isolated 
extrema of $f$ inside that closed surface, must satisfy the  relation (\ref{poincare}). 
This condition is satisfied on general grounds because extreme compression or stretching of a bond makes the energy very large compared to the equilibrium energy.
There have only been a few previous discussions about how global constraints such
as equation (\ref{poincare}) follow from differential topology.
Mezey considered lower and upper bounds for $N_{min}$, $N_{max}$, and $N_s$,
based on the Morse inequalities \cite{Mezey:1981,Mezey:1987}. 
Pick considered the corresponding equation [which has
zero on the right hand side] for absorbates on periodic substrates \cite{Pick:2009}.

A study of a simple model two-dimensional potential for double proton transfer \cite{Benderskii:2003}
also produced a phase diagram as a function of the model parameters and considered
bifurcation of the instanton tunneling paths. However, it has been argued that the model potential used there does
not respect some of the symmetries of the problem \cite{Smedarchina:2007}.

\subsection{Co-operativity of hydrogen bonds}

The ground state energy of the double H-bond system can be compared to that
of two decoupled H-bonds where $D_1$ is the binding energy of an isolated X-H bond. The energy of a single H-bond for the corresponding two-diabatic state model \cite{McKenzieCPL} is
\begin{equation}
\epsilon_0^1(R,r)={1 \over 2}\left(V_1(r)+V_1(R-r)-\sqrt{(V_1(r)-V_1(R-r))^2 + 4\Delta(R)^2} \right)
\label{eqn-single}
\end{equation}
and the ground state energy of the two decoupled H-bonds is
\begin{equation}
\epsilon_0(R,r_1,r_2)=\epsilon_0^1(R,r_1)+\epsilon_0^1(R,r_2).
\label{eqn-double}
\end{equation}
It turns out that for $D_2=D_1$ the ground state energy of the Hamiltonian (\ref{eqn-ham})
is given by an identical expression.
Thus, we see that it is differences between the single and double proton
affinities that couple the two H-bonds together.
The question of whether the two bonds are co-operative or not is subtle.
Is the binding energy enhanced or decreased? Is the barrier for double proton transfer increased or decreased?
Figure 4 shows that as $D_2/D_1$ decreases (and thus the coupling between the two H-bonds
increases) that the $R$ value needed to remove the barrier decreases. This is a signature of
anti-co-operative behaviour. Similarly the barrier for concerted proton transfer
increases as $D_2/D_1$ decreases.

\section{Model parameter values for specific molecules}

The above analysis shows that a key parameter in the model is  $D_2/D_1$,
the ratio of the
second proton affinity, $D_2$ 
to 
the first proton affinity, $D_1$.
Unfortunately, there are few calculations or measurements of
$D_2$ for specific molecules, that I am aware of.
DFT-B3LYP and MP2 calculations for diazanaphthalenes give
$D_2/D_1 \simeq 0.5$ \cite{Spedaletti:2005}. DFT calculations for a series of diamines containing hydrocarbon
bridges give $D_2/D_1 \sim 0.8-0.9$ \cite{Bayat:2010}.
For uracil, DFT-B3LYP calculations give a proton affinity of 800-900 kJ/mol depending 
on the protonation site and the tautomer, and a deprotonation enthalpy of 
1350-1450 kJ/mol \cite{Kryachko}.
This is equivalent to $D_2/D_1 \simeq 0.6-0.7$.
For the chromophore of the Green fluorescent protein, the ground state energies of four different protonation states have been calculated at the level of MS-CASPT2 with a SA2-CAS(4,3) reference at the MP2 ground state geometry with a cc-pvdz 
basis set \cite{Olsen:2010}. With respect to the anion state the energies of the phenol, imidazolol, and oxonol cation states
are -15.09, -13.74, and -24.72 eV, respectively.
This leads to $D_2/D_1 \simeq 0.7-0.8$.
Given the values discussed above, the parameter range $0.5 < D_2/D_1 < 1$ shown in Figure 4 is chemically realistic.
 
For porphycenes chemical substitutions external
to the cavity produce a range of $R$ values for the N-H$\cdots$N bonds inside
the cavity 
between 2.5 and 2.9 $\AA$ \cite{Waluk:2010}.
This variation leads to the tautomerisation [double proton transfer] rate increasing smoothly by more than three orders of magnitude as $R$ decreases \cite{Fita:2009}.
This is consistent with the fact that in the model presented here
the scale of the energy barriers is largely
determined by $V(R/2)-\Delta(R)$ which decreases with decreasing $R$.

The plots shown in this paper use a parameterisation for O-H$\cdots$O bonds \cite{McKenzieCPL}.
The parameterisation will be slightly different for N-H$\cdots$N  bonds.
For N-H bonds  Warshel \cite{Warshel:1991} has Morse potential
parameter values: $r_0 = 1.00 \AA$, $a=2.07/\AA$, and 
$D =103$ kcal/mol, compared to values for  O-H of
$r_0 = 0.96 \AA$, $a=2.26/\AA$, and 
$D =102$ kcal/mol.
These differences will increase the $R$ values on the vertical scale of Figure \ref{figg} by approximately $0.14 \AA$, assuming the same parameterisation of $\Delta(R)$,
i.e., $b=a$ and $\Delta_1=0.4 D$.

For single hydrogen bonds, with symmetric donor and acceptor,  empirical correlations between $R$ and a range
of observables such as bond lengths, vibrational frequencies, NMR chemical shifts, and geometric isotope effects are observed \cite{Gilli,McKenzieCPL,McKenzieJCP}. Our model suggests that for systems with double hydrogen bonds such
correlations will only occur for systems with identical values of the parameter $D_2/D_1$.
For carboxylic acid dimers it was found that $R$ coud be varied systematically by enclosing them in
a molecular capsule \cite{Ajami:2011}.
 Families of porphycenes \cite{Waluk:2010} may also be appropriate systems to investigate
such correlations.

An important task is to determine in which of the four classes
specific molecular systems belong.
The formic acid dimer is the simplest dimer of carboxylic acids.
A tunnel splitting of $2.86 \times 10^{-3}$ cm$^{-1}$ for the C=O stretch mode
has been observed at low temperatures \cite{Madeja:2002} and identified with
a tunneling path associated with concerted transfer.
A transition from class II to III with increasing $R$ was observed in 
quantum chemistry calculations for the formic acid dimer.
Shida et al. \cite{Shida:1991} found a transition from one to  two saddle points
when $R \simeq 2.7 \AA$. 
Smedarchina et al.\cite{Smedarchina:2008}
found a transition for $R=2.56 \AA$.
In the simple model considered hear the transition occurs for $R \simeq 2.4 \AA$.
 The experimental value for the equilibrium bond length
is $R=2.696 \AA$ from electron diffraction, compared to the value of 
$R=2.72 \AA$ based on DFT calculations \cite{Smedarchina:2008}
suggesting that formic acid is in class III.

An interesting question is whether there are any compounds that fall in
class I, i.e., with completely delocalised protons.
For N-H$\cdots$H bonds this will require $R < 2.35 \AA$, which is fairly unlikely.
However, the quantum zero-point motion of the protons may
lead to delocalisation at larger distances, $R \simeq 2.5 \AA$, provided
that the energy barrier is less than the zero-point vibrational energy \cite{McKenzieJCP}.










\section{Conclusions}

A simple model has been introduced that can describe 
four different classes of potential energy
surfaces for double proton transfer in symmetric hydrogen-bonded complexes, such as porphycenes and carboxylic acid dimers.
The model is based on a chemically and physically transparent effective Hamiltonian involving four diabatic states.
The number of saddle points on the ground state potential energy surface is determined by the value of two different parameters: $R$, the distance between the donor and acceptor atoms
for proton transfer, 
and $D_2/D_1$, the ratio of the proton affinity of a donor with two and one protons attached.
Double proton transfer will occur via a concerted, partially concerted, or sequential mechanism depending
on the class of potential energy surface involved.

\begin{acknowledgments}

I thank Seth Olsen, David Wales, and Kjartan Thor Wikfeldt  for helpful discussions. 

\end{acknowledgments}

\end{document}